\documentclass[superscriptaddress,prb,twocolumn,showpacs,amsmath,amssymb,letterpaper]{revtex4}

\usepackage{bm}% bold math

\usepackage{graphicx}
\usepackage{latexsym}
\usepackage{amsmath}
\usepackage{amssymb}
\usepackage{amsfonts}
\usepackage{color}
\usepackage{bm}% bold math
\usepackage{verbatim}

\begin{document}

\title{Microscopic theory of type-1.5 superconductivity in two-band systems}

\author{Mihail Silaev}
\affiliation{Department of Theoretical Physics, The Royal Institute of Technology, Stockholm, SE-10691 Sweden}
\affiliation{Institute for Physics of Microstructures RAS, 603950 Nizhny Novgorod, Russia.}
\affiliation{ Department  of Physics, University of Massachusetts Amherst, MA 01003 USA }
\author{Egor Babaev}
\affiliation{Department of Theoretical Physics, The Royal Institute of Technology, Stockholm, SE-10691 Sweden}
\affiliation{ Department  of Physics, University of Massachusetts Amherst, MA 01003 USA }

\date{\today}

%%% abstract
\begin{abstract}
%Using a self-consistent microscopic theory, we  prove, at a microscopic level, that
%``type-1.5" superconductivity is possible in multi-band systems,
%despite interband coupling breaking the symmetry down to $U(1)$.
%First we report a microscopic theory of the type-1.5 behavior at
%elevated temperatures which is consistent the predictions of
%phenomenological GL models. In the second part of the paper
We report a self-consistent  microscopic theory  of
characteristic length scales, vortex structure and
  type-1.5 superconducting state in two-band systems
  using two-band Eilenberger formalism.
 \end{abstract}

%%% PACS numbers
\pacs{74.25.QP, 74.25.Fy, 73.40.Gk}
\maketitle

\section{Introduction}
The usual classification of  superconductors
 characterizes materials by the Ginzburg-Landau parameter $\kappa$
(which is the ratio   of
the characteristic length scale of the order parameter variation $\xi$
and the magnetic field penetration length $\lambda$) \cite{GL}.
The remarkable property is that
within the GL theory of single-component superconductivity $\kappa$ determines
 the major features of the phase diagram of the system
in magnetic field.
In type-I superconductor  $\kappa<1/\sqrt{2}$  (i.e.  order parameter
is the slowest varying field), %the energy of superconductor/normal metal boundary is positive,  besides that
 vortex excitations have attractive interaction and  are thermodynamically
unstable in applied magnetic field. %(i.e. $H_{c1}$ is smaller than the thermodynamical magnetic field).
Thus in an applied field a type-I system forms macroscopically large normal domains \cite{deGennes}.
For $\kappa>1/\sqrt{2}$  (type-II superconductivity)
% the energy of superconducting/normal boundary is negative,
vortices are
thermodynamically stable and interact repulsively yielding a new
 phase in strong magnetic fields: a lattice of quantized vortices
\cite{abrikosov,deGennes}. In the Bogomolnyi limit
($\kappa=1/\sqrt{2}$) the vortices do not interact in the
Ginzburg-Landau theory. However indeed it should be remarked that
going to a deeper microscopic level there are always
``next-to-leading order" microscopic corrections.
 These corrections, though unimportant even slightly away from this limit,
provide weak non-universal intervortex interactions when $\kappa$
is very close to $1/\sqrt{2}$
  see e.g. \cite{Klein,Jacobs-VortexAttraction}.
Apparently a counterpart of this limit  is also   possible in
multi-component systems. { However in this case the Bogomolnyi
limit could appear only via quite extreme fine-tuning of
parameters and therefore is not of much physical relevance.}  In
this work we are interested only in the entirely different physics
of intervortex interactions and magnetic response of
multicomponent systems originating from the different funamental
length scales
 { very far} from any counterparts of Bogomolnyi limit.

A question which attracted much attention recently is whether the
type-I/type-II classification is sufficient for characterizing the
rapidly growing family of multicomponent systems of physical
interest \cite{bs1}. A clear cut  example of the system where
type-I/type-II dichotomy does not hold is  the projected
coexistent electronic and protonic (or deuteronic)
superconductivity \cite{nature} in hydrogen isotopes, their
mixtures and hydrogen rich alloys at ultrahigh compression as well
as the coexisting protonic and
$\Sigma^-$-hyperonic superconductivity in  neutron stars. %\cite{neutr}.
 These systems have  $U(1) \times U(1)$  or higher symmetries
and thus several   fundamental length scales associated with
independently conserved fields. Consequently the system cannot be
characterized by a single dimensionless parameter $\kappa$. In an
applied field the only thermodynamically stable vortex solutions
are ``composite" vortices  where both condensates have $2\pi$
phase windings. Consequently such vortices have cores in both
components \cite{frac,nature}. Importantly it also acquires a new
regime \cite{bs1} for which the term ``type-1.5" was coined
recently \cite{m1}. In that regime like in a type-I case the
characteristic core size of one of the components is larger than
the flux carrying area. The overlap of these   cores produces
attractive intervortex interaction. However, in contrast to type-I
case, these vortices  have repulsive interaction at short ranges.
\cite{bs1,GL2mass,GL2mass2,semimeissner}.
 This kind of non-monotonic vortex interaction results in the
appearance of the additional ``semi-Meissner" phase in low
magnetic fields. In that phase vortices  form clusters where
because of overlap of cores the slowest varying density component
is suppressed. Moreover these vortex clusters coexist with the
domains of two-component Meissner state.

The recent experimental works proposed that two-band  \cite{Suhl}
electronic material $MgB_2$ belongs to the type-1.5 case
\cite{m1,m2}. The principal difference with the discussed above
$U(1)\times U(1)$ theory is that interband coupling breaks the
symmetry down to $U(1)$ (for a recent discussion of microscopic
details see e.g. \cite{gurevich,agterberg}). Therefore there is a
single superconducting phase transition at a single $T_c$.
 However, at the same time the system has two gaps
and two superfluid densities, which, in general vary at distinct
characteristic length scales at { any} finite distance from $T_c$.
Therefore the type-1.5 magnetic response can arise even
infinitesimally far away from $T_c$ from the interplay of  two
density modes which originate from the underlying two-gap physics.
This behaviour  was demonstrated in the framework of
phenomenological two-component $U(1)$ GL
 models \cite{GL2mass,GL2mass2}.
 %(for a review of  derivations of such GL models see \cite{gurevich}).
%However, indeed, a multiband superconductor does not necessary
%permit an expansion yielding a type-1.5 two-component GL field
%theory.
 %Thus  to study type-1.5
%superconductivity in  multiband systems in general case one
%should resort
Here we develop a theory of type-1.5 superconductivity
based on a microscopic theory without
involving a GL expansion.
That is, in this work  we use the Eilenberger formalism and  demonstrate the existence
as well as describe basic properties  of type-1.5 superconductivity in multiband
materials.

\section{Microscopic description of vortex state in multiband superconductor}

\subsection{Eilenberger formalism}

We consider a superconductor with two overlapping bands at the
Fermi level  \cite{Suhl}. The corresponding two sheets of the
Fermi surface are assumed to be cylindrical. Within quasiclassical
approximation the band parameters characterizing the two different
sheets of the Fermi surface are the Fermi velocities $V_{Fj}$ and
the partial densities of states (DOS) $\nu_j$, labelled by the
band index $j=1,2$. We normalize the energies to the critical
temperature $T_c$ and length to $r_0= \hbar V_{F1}/T_c$. The
system of Eilenberger equations for two bands is
\begin{align}\label{Eq:EilenbergerF}
&v_{Fj}{\bf n_p}\left(\nabla+i {\bf A}\right) f_j +
 2\omega_n f_j - 2 \Delta_j g_j=0, \\ \nonumber
 &v_{Fj}{\bf n_p}\left(\nabla-i {\bf A}\right) f^+_j -
 2\omega_n f^+_j + 2\Delta^*_j g_j=0.
 \end{align}
 Here $\omega_n=(2n+1)\pi T$ are Matsubara frequencies and
  $v_{Fj}=V_{Fj}/V_{F1}$. The vector ${\bf n_p}=(\cos\theta_p,\sin\theta_p)$
  parameterizes the position on 2D cylindrical
 Fermi surfaces. The quasiclassical Green's functions in each band obey
 normalization condition $g_j^2+f_jf_j^+=1$.
 The self-consistency equation for the gaps is
 \begin{equation}\label{Eq:SelfConsistencyOP}
  \Delta_i=T \sum_{n=0}^{N_d} \int_0^{2\pi}
 \lambda_{ij} f_j d\theta_p.
 \end{equation}
 The coupling matrix $\lambda_{ij}$ satisfies the symmetry relations
 $n_1\lambda_{12}=n_2\lambda_{21}$ where $n_i$ are the
 partial DOS normalized so that $n_1+n_2=1$.
We consider $\lambda_{11}>\lambda_{22}$ and
 therefore refer to the first band as ``strong" and to the second
 as ``weak". The vector potential
  satisfies the Maxwell equation
 \begin{equation}\label{Eq:Maxwell}
 \nabla\times\nabla\times {\bf A} = {\bf j}
 \end{equation}
  where the current is
  \begin{equation}\label{Eq:SelfConsistencyCurrent}
   {\bf j}= -T\sum_{j=1,2} \sigma_j\sum_{n=0}^{N_d}
 Im\int_0^{2\pi}  {\bf n_p} g_j d\theta_p.
\end{equation}
 The parameters $\sigma_j$ are given by
 $$
 \sigma_j=\pi\left(\frac{4e}{c}\right)^2 (r_0 V_{F1})^2
 \nu_j v_{Fj}.
 $$

 \subsection{Multiple masses of the $\Delta$ fields}

 First we focus on the structure of an isolated axially symmetric vortex
 characterized by the non-trivial phase winding of
 the gap functions $\Delta_{1,2}=|\Delta_{1,2}|(r)e^{i\varphi}$.
 We begin by finding the asymptotics of the gap
 function modules $|\Delta_{1,2}|(r)$ at distances far from the vortex core. In this case the
 Eilenberger Eqs.(\ref{Eq:EilenbergerF}) can be linearized by generalizing the methods used
for single band superconductors \cite{Leung-Jacobs}. The details
of the asymptotics derivation are given in the
Appendix\ref{Appendix1}. We rewrite the
Eqs.(\ref{Eq:EilenbergerF}) in terms of the deviations from the
vacuum state values
 $\bar{\Delta}_j=\Delta_{j0}-|\Delta_j|$ and
 $\bar{f}_j={f}_{j0}-{f}_j$,
  $\bar{f}^{+}_j={f}^{+}_{j0}-{f}^{+}_j$
 keeping on the left side the first order terms.
Then we take the real part of the Eqs.(\ref{Eq:EilenbergerF})
 to obtain the following system
 \begin{align}\label{Eq:EilenbergerF-lin-R}
 &v_{Fj}{\bf n_p}\nabla \bar{f}^r_{\Sigma j}+2\omega_n
 \bar{f}^r_{dj}=X^r_{\Sigma j} \\ \nonumber
 & v_{Fj}{\bf n_p} \nabla \bar{f}^r_{dj}+2\frac{\Omega_n^2}{\omega_n} \bar{f}^r_{\Sigma
 j}-
 \frac{4\omega_n}{\Omega_{nj}}\bar{\Delta}_j=X^r_{dj},
 \end{align}
 where $\Omega_{nj}=\sqrt{\omega_n^2+\Delta_{0j}^2}$,
   $\bar{f}^r_{\Sigma j}=Re [\bar{f}_j+\bar{f}^+_j]$
  and $\bar{f}^r_{d j}=Re [\bar{f}_j-\bar{f}^+_j]$. In Eqs.(\ref{Eq:EilenbergerF-lin-R})
  the higher order terms in $\bar{\Delta}_j$, $\bar{f}$ and $\bar{f}^+$ are incorporated in the
 right hand side (r.h.s) source functions $X_{\Sigma (d) j}=X_{\Sigma (d) j}({\bf n_p},\omega_n, {\bf
 r})$.

The solution of Eqs.(\ref{Eq:EilenbergerF-lin-R}) can be found in
the momentum representation $f^r_{\Sigma (d)j} ({\bf k})=\int
f^r_{\Sigma (d) j}({\bf r}) \exp (-i {\bf k r}) d^2 {\bf r}$.
After substituting it to the self-consistency equation we get the
expression for the { gap functions}
 \begin{equation}\label{Eq:OpFourier}
 \bar{\Delta}_i(k)= \hat R^{-1}_{ij} N_j(k).
 \end{equation}
 The elements of the matrix $\hat R=\hat R(k)$ are
 $R_{ii}=(\lambda_{ii} S_i-1)$ and $R_{ij}=\lambda_{ij} S_j$, where
\begin{equation}\label{Eq:Sj}
 S_j(k)=4\pi T\sum_{n=0}^{N_d}
 \frac{\omega_n^2}{\Omega_{nj}^2}\left[4\Omega_{nj}^2+(v_{Fj}
 k)^2\right]^{-1/2}.
\end{equation}
  The source functions $N_j(k)$ come from
the r.h.s of Eqs.(\ref{Eq:EilenbergerF-lin-R}). The strict
definition of source functions is given in the
Appendix\ref{Appendix1}.

The real space asymptotic of the gap functions
(\ref{Eq:OpFourier}) is determined by the contributions of the
singularities of the response function
 $\hat R^{-1}(k)$ which are poles at the zeros of the determinant
 $D_R(k)={\rm Det} [\hat R(k)]$ and branch points at $k=2i\Omega_{nj}/v_{Fj}$.
 Similarly to  Ref.(\onlinecite{Leung-Jacobs}) we assume the branch
 cuts to lie along the imaginary axis from $k=2i\Omega_{nj}/v_{Fj}$ to
 $k=i\infty$.
To find the asymptotics of the gaps $\bar{\Delta}_i(r)$, we need
only to take into account the poles of
 $\hat R^{-1}(k)$ lying in the upper complex half plane below all the
 branch cuts. In this case all the
 zeros of the function $D_R (k)$ are purely imaginary $k=i\mu_n$.
 Each of them can be associated with the particular
 mass $\mu_n$ of the {\it composite mode} formed by a {\it superposition} of
 gap functions in two superconducting bands.
 The composite character of the modes arises in our case because
 the two bands are directly coupled.
 The inverse of the mass controls the characteristic  length scale
at which this superposition of the gap fields varies. Therefore the
lightest mass determines very-long-distance decay of {\it both}
$\bar{\Delta}_1$ {\it and} $\bar{\Delta}_2$.
  The contribution from the branch cut contains all the length scales which are smaller than
 the threshold one given by position of the lowest branch point
 $k=i q_{bp}$ where $q_{bp}=2\min(\Omega_{02}/v_{F2},
 \Omega_{01}/v_{F1})$.

The Eq.(\ref{Eq:OpFourier}) results in the asymptotical expression
for the gap functions
 $$
 \bar{\Delta}_i(r)=\int_0^r dr_1 G_{ij} (r,r_1) N_j(r_1).
 $$
 Here $N_j(r)$ is the Fourier-Bessel image of
the source function in Eq.(\ref{Eq:OpFourier}) and
\begin{align}
  &\hat G(r,r_1)=\sum_n \hat A_n K_0(q_n r)I_0(q_n r_1)+ \\ \nonumber
  &\frac{2}{\pi}\int_{q_{bp}}^{\infty} ds s K_0(sr)I_0(sr_1)
\left[\hat R^{-1}\right]_{k=is} \nonumber
\end{align}
 where $K_0$ and $I_0$ are
MacDonald and modified Bessel functions. The matrices $\hat A_n$
determining the contributions of the pole terms are
\begin{equation}\label{Eq:MatrixA}
\hat A_n =2ik\left[\frac{dD_R}{dk}\right]^{-1} \begin{pmatrix}
   R_{22} & -R_{12} \\
   -R_{21} & R_{11} \
 \end{pmatrix}|_{k=iq_n}.
\end{equation}
 and the branch cut contribution is determined by the
 the jump of the response function
 \begin{equation}
\left[\hat R^{-1}\right]_{k=is}=\hat R^{-1}(k=is+0)-\hat R^{-1}(k=is-0).
\end{equation}
 Under rather general conditions, the response function in Eq.(\ref{Eq:OpFourier}) has
 two poles given by zeros of the determinant $D_R(k)=0$ which lie
below the branch cuts. Thus in this case the asymptotical
behaviour of the gap functions is principally different from the
single band superconductor, despite the fact they share the same
$U(1)$ symmetry of the order parameter. The two poles determine
the {\it two inverse length scales} or, equivalently, the two
masses of composite gap functions fields, which we denote as
``heavy" $1/\xi_H=\mu_H$ and ``light" $1/\xi_L=\mu_L$ (i.e.
$\mu_H>\mu_L$).
 The corresponding composite gap function modes are
parameterized by the two ``mixing angles" $\theta_L,\theta_H$ as
follows:
\begin{equation}\label{Eq:MixingAngle}
\left(\tilde{\Delta}_L\atop \tilde{\Delta}_H \right)=
 \begin{pmatrix}
   \cos\theta_L & \sin\theta_L \\
   -\sin\theta_H & \cos\theta_H \
 \end{pmatrix}
 \left(\bar{\Delta}_1\atop \bar{\Delta}_2\right).
  \end{equation}
 Note that in the two-band  GL theory
without interband impurity scattering terms one has
$\theta_L=\theta_H$ \cite{GL2mass,GL2mass2}. Below we recover this
behavior at elevated temperatures without using GL-like expansion,
thereby verifying predictions of phenomenological GL models.
However,   outside the range of validity of the GL theory we find
that $\theta_L \ne \theta_H$.

  Let us now consider in detail an example of the system with
 $\lambda_{11}=0.25$, $\lambda_{22}=0.213$, $n_1=n_2=0.5$
 and various values of the interband coupling $\lambda_J=\lambda_{12}=\lambda_{21}$.
 We focus on the two different regimes, determined by the band parameter
 $\gamma_F=v_{F2}/v_{F1}$ namely
 {\bf (i)} $\gamma_F>1$ and {\bf (ii)} $\gamma_F<1$.

\begin{figure}[h!]
\centerline{\includegraphics[width=1.0\linewidth]{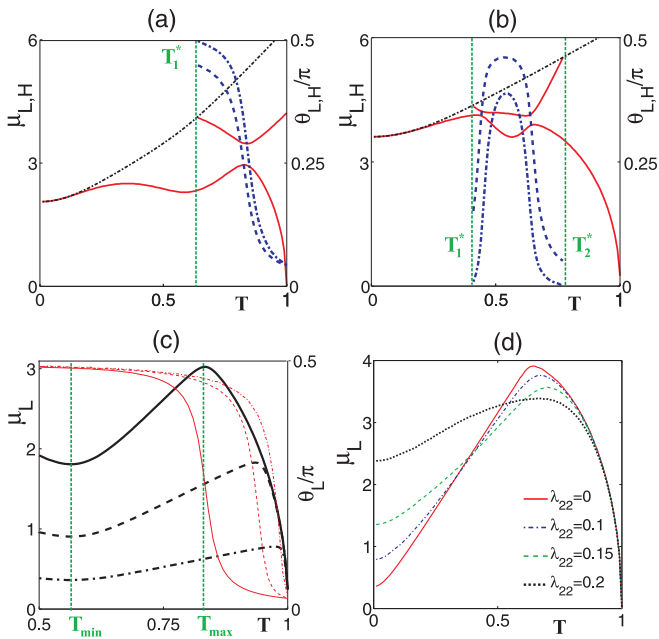}}
\caption{\label{Fig:Asymptotics1} Masses $\mu_{L,H}$ of the
composite gap function fields for (a) $\gamma_F=1$ and
$\lambda_{J}=0.005$, (b) $\gamma_F=0.5$ and $\lambda_J=0.0025$.
The position of branch cut is shown by black dashed line. The
mixing angles $\theta_{L,H}$ are shown by blue dashed and
dash-dotted lines correspondingly. (c) Temperature dependence of
the mass $\mu_L (T)$ (black thich curves) and the corresponding
mixing angle $\theta_L$ determined by Eq.(\ref{Eq:MixingAngle})
(red thin curves) for $\gamma_F=1;\;2;\;5$ (solid, dashed and
dash-dotted curves). The coupling parameters are
$\lambda_{11}=0.25$, $\lambda_{22}=0.213$ and $\lambda_J=0.005$.
(d) Temperature dependence of the mass $\mu_L (T)$ for different
values of coupling constant $\lambda_{22}$.  }
\end{figure}

\begin{figure}
\centerline{\includegraphics[width=1.0\linewidth]{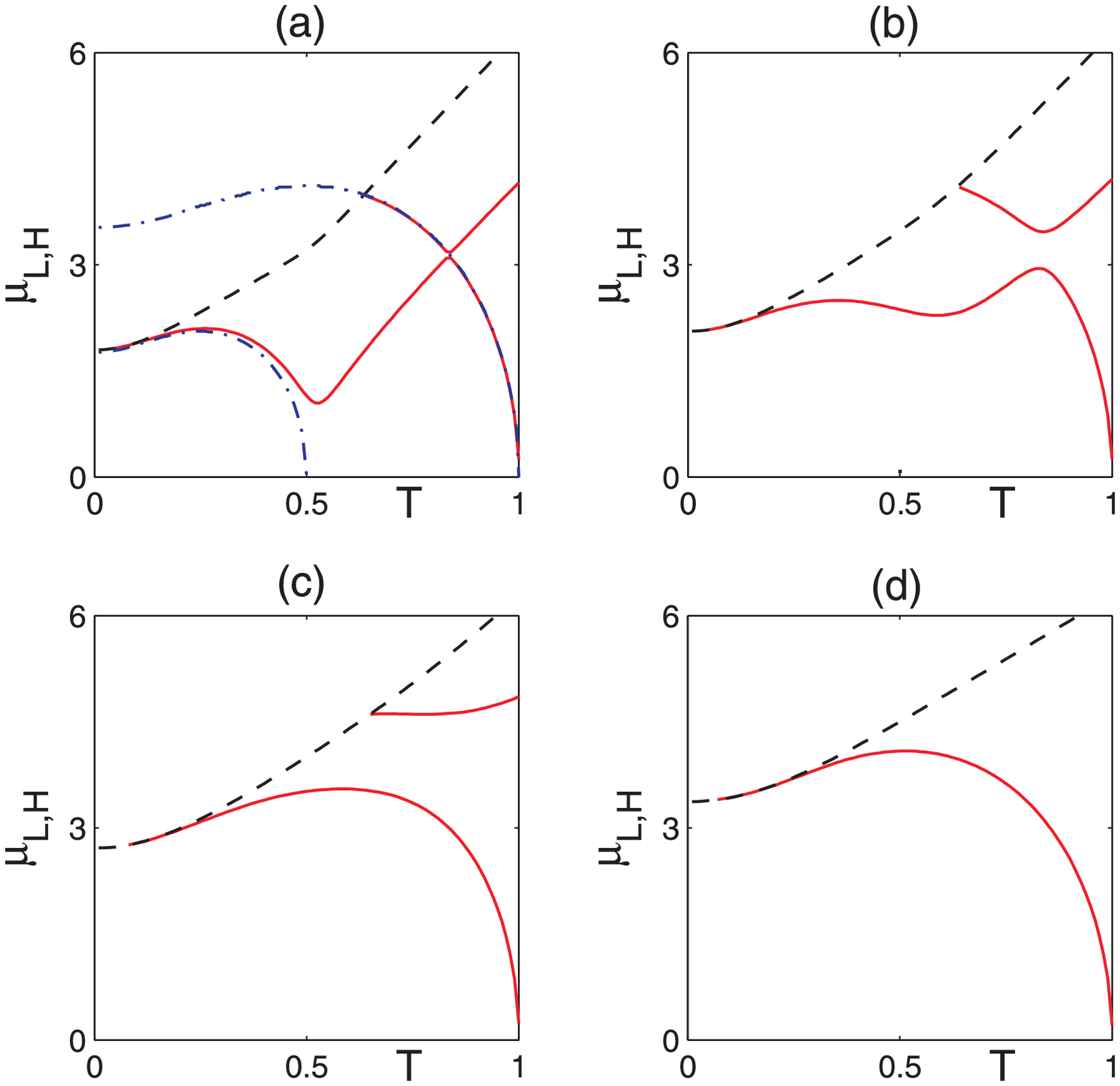}}
\caption{\label{Fig:SequenceModes} Masses $\mu_{L}$ and $\mu_{H}$
(red solid lines) of the composite gap function fields for the
different values of interband Josephson coupling $\lambda_J$ and
$\gamma_F=1$. In the sequence of plots (a)-(d) the transformation
of masses is shown for $\lambda_J$ increasing from the small
values $\lambda_J\ll \lambda_{11},\lambda_{22}$ to the values
comparable to intraband coupling $\lambda_J\sim
\lambda_{11},\lambda_{22}$. The particular values of coupling
constants are $\lambda_{11}=0.25$, $\lambda_{22}=0.213$ and
$\lambda_J=0.0005;\;0.0025;\;0.025;\;\lambda_{22}$ for plots (a-d)
correspondingly. The branch cuts are shown by black dashed lines.
 In (a) with blue
dash-dotted lines the masses of modes are shown for the case of
$\lambda_J=0$. Note that at $\lambda_J=0$ the two masses go to
zero at two different temperatures. Because $1/\mu_{L,H}$ are
related to the coherence lengths, this reflects the fact that for
$U(1)\times U(1)$ theory there are two independently diverging
coherence lengths. Note that for finite values of interband
coupling only one mass $\mu_L$ goes to zero at one $T_c$.
  }
\end{figure}

 {\bf (i)} Some of the basic properties of this regime are captured by the particular
 case when $\gamma_F=1$. The examples of the temperature dependencies
 of the masses $\mu_{L,H}(T)$ are shown in the Fig.\ref{Fig:Asymptotics1}(a).
 The two  massive modes coexist at the temperature interval
 $T^*_1<T<T_c$, where the temperature $T^*_1$
 is determined by the branch cut position, shown in the
 Fig.\ref{Fig:Asymptotics1}(a) by black dashed line. For temperatures
 $T<T^*_1$ there exists only one massive mode. At very low temperatures the mass $\mu_L$
 is very close to the branch cut. As the interband coupling
 parameter is increased, the temperature $T^*_1$ rises and becomes equal to $T_c$
 at some critical value of $\lambda_J=\lambda_{Jc}$. For the particular case
 of $\gamma_{F}=1$ we found an exact condition
 $\lambda_{Jc}=\lambda_{22}$.  The evolution of the masses $\mu_{L,H}$ is shown in the sequence of plots
 Fig.\ref{Fig:SequenceModes}(a)-(d)  for $\lambda_J$ increasing from the small
values $\lambda_J\ll \lambda_{11},\lambda_{22}$ to the values
comparable to intraband coupling $\lambda_J\sim
\lambda_{11},\lambda_{22}$.

 {\bf (ii)} In the case if $\gamma_{F}<\gamma_{th}$ (where
 $\gamma_{th}$ is a characteristic value determined by the system
 parameters) the two massive modes
 coexist at some temperature interval $T^*_2<T<T^*_1$ where $T^*_1\leq T_c$.
 For the particular case when $\gamma_F=0.5$,
 the temperature dependencies of $\mu_{L,H}(T)$ are shown in the
 Fig.\ref{Fig:Asymptotics1}(b).

In Fig.\ref{Fig:Asymptotics1}(a,b) the mixing angles $\theta_L$
and $\theta_H$ given by Eq.(\ref{Eq:MixingAngle}) are shown by
blue dashed and dash-dotted lines correspondingly. In the case
{\bf (i)} near the critical temperature the angles are
approximately equal, which  provides  for this regime a
microscopic verification for  of the results obtained using
phenomenological GL theories \cite{GL2mass,GL2mass2}. At lower
temperatures the discrepancy is considerable and grows with the
increasing interband coupling. Large deviations of the  mixing
angle from $0$ and $\pi/2$ signal strong mixing of the gap fields.
It occurs near the avoided crossing points of $\mu_{L} (T)$ and
$\mu_{H} (T)$. In case {\bf (i)} shown in
Fig.\ref{Fig:Asymptotics1}(a) there is   one avoided crossing
point and in the case {\bf (ii)} in   there can be two of them, as
shown in Fig.\ref{Fig:Asymptotics1}(b).

%%%%%%%%%%%%% FIGURE %%%%%%%%%%%%%%%%%%

%%%%%%%%%%%% FIGURE %%%%%%%%%%%%%%%%%%

 The discussed above existence of two modes
associated with mixed gap functions { can, under certain conditions, result in the
type-1.5 behavior as it was demonstrated in the framework of GL
approach\cite{GL2mass,GL2mass2}}. However, importantly the microscopic
formalism we use here allows to describe type-1.5
superconductivity beyond the validity of GL models. The type-1.5
behavior requires a density mode with low mass $\mu_L$ to mediate
intervortex attraction at large separations, which should coexist
with short-range repulsion.

 We find that the temperature dependence of
$\mu_L(T)$ is characterized by an anomalous behavior, which is in
strong contrast to temperature dependence of the mass of the  gap
mode in single-band theories. As shown on
Fig.\ref{Fig:Asymptotics1}(c) the function $\mu_L(T)$ is {\it
non-monotonic} with the minimum at the temperature $T_{min}$. The
minimum is close to the crossover temperature
 where the second superconducting band becomes
active. The maximum is located at the temperature
$T_{min}<T_{max}<T_c$.

The structure of the composite gap function mode shown in the
Fig.\ref{Fig:Asymptotics1}(c) $\tilde{\Delta}_L$ is characterized
by the mixing angle $\theta_L$ given by Eq.(\ref{Eq:MixingAngle}).
At the temperature interval $T<T_{max}$ the mixing angle is
$\theta_L\approx \pi/2$.
  Therefore in this temperature regime, the mode with lightest mass consists
primarily of the weak band gap $\bar{\Delta}_2(r)$ with a tiny
admixture of $\bar{\Delta}_1(r)$. Note that in this regime the
overall behavior of $|\Delta_1|(r)$ outside the long-range
asymptotic tail has relatively weak dependence on interband
coupling (i.e. at larger distances from the core it has slowly
recovering tail associated with only tiny suppression relative to
its ground state value).
 At the same time the recovery of
 $|\Delta_2|(r)$ to a larger degree is  dominated
by the light mass mode.

\subsection{The high temperature limit.}

 As noted above at elevated temperatures the mixing angles
have close values, consistently with the type-1.5 behaviour which
appears in the framework of  two-band Ginzburg-Landau models
\cite{GL2mass2}. At very high  temperatures $T_{max}<<T<T_c$ the
mixing angle $\theta_L$ gradually becomes small $\theta_L\ll \pi$,
which means that there the mode $\tilde{\Delta}_L$ is dominated by
the strong band contribution $\bar{\Delta}_1$.

 Since any Josephson interband coupling breaks the symmetry of
the system in question down to $U(1)$, then according to
Ginzburg-Landau argument this symmetry dictates that,
asymptotically, {\it in the limit} $T \to T_c$ one should recover
a single-component-like GL temperature dependence $\mu_L\sim
\sqrt{1-T/T_c}$ of a single order parameter (at the level of
mean-field theory) \cite{GL}.

In the regimes corresponding to Fig.\ref{Fig:Asymptotics1}(a,c)
very close to $T_c$ the mixing angle of the heavy mode is small
$\theta_H\ll 1$ which makes the contribution of the smaller gap
$\bar{\Delta}_2$ to the heavy mode the dominating one. This
behaviour of the mixing angles, and the fact that that for
non-zero Josephson coupling only one mass $\mu_L(T)$ goes to zero
at $T\to T_c$ allows one to neglect the { heavy mode} and
construct a mean-field GL order parameter with the scaling
$\mu_L\sim \sqrt{1-T/T_c}$ {\it as an ``asymptotic"
characteristic} in the limit $T \to T_c$. However as shown in the
Fig.\ref{Fig:Asymptotics1}(c) the temperature region of such
behavior shrinks drastically for large disparities of the band
characteristics and weak interband couplings. In general the
smaller is the interband coupling, the closer to $T_c$ one should
be in order to obtain single component-like GL scaling. For a wide
range of parameters the mean field GL theory with the single
component-like scaling $\mu_L\sim \sqrt{1-T/T_c}$ will emerge only
infinitesimally close to $T_c$.Note that the limit where
$\mu_L\sim \sqrt{1-T/T_c}$ is in certain cases unphysical because
the underlying mean-field theory can become invalid because of
fluctuations, at temperatures lower that the temperature where
this scaling would take place. Thus even in weak-coupling two-band
systems with $U(1)$ symmetry, for a wide parameter range, one
could not apply a leading order in ${(1-T/T_c)}$ GL theory since
the region of its applicability will fall into the parameter space
where underlying mean field theory is not valid because of
fluctuations.
 In contrast to single-component systems, as the
consequence of the presence of two gaps even slightly away from
$T_c$ the behaviour of $\mu_L(T)$ can be drastically different
from the usual GL scaling. As a result the product $\Lambda\mu_L$
where $\Lambda$ is the magnetic field penetration length acquires
a strong temperature dependence. Moreover as we  show below, its
limiting value at $T_c$ does not determine entirely the
intervortex interaction potential nor the magnetic response of the
system. Therefore one cannot in general parameterize  the magnetic
response of two-band systems by the single GL parameter
$\kappa={\Lambda}/{\xi}$.

\subsection{Light mode of { gap function field} and type-1.5 behavior.}

The plots of $\mu_L(T)$ for $\gamma_F=1;2;5$ are shown in
Fig.\ref{Fig:Asymptotics1}(c) by solid, dashed and dash-dotted
thick black lines. There is a clear general tendency of increasing
$T_{max}$ with growing  parameter $\gamma_F$ which characterizes
band disparity. It leads to broadening of the temperature region
of the anomalous behavior of the mass $\mu_{L}(T)$   where the
fields asymptotics are dominated by the weak band.
 The Fig.\ref{Fig:Asymptotics1}(c) clearly demonstrates the considerable overall
suppression of  $\mu_L$ with growing parameter $\gamma_F$. {The
inverse of  the mass of the light composite gap mode $\mu_L$ sets
the range of the attractive density-density contribution to
intervortex interaction. Therefore the condition for the
occurrence of the intervortex attraction will be met if $\mu_L$ is
smaller than  $\Lambda^{-1}$.

Thus a physically important situation arising in  a two-band
superconductor, is that  { for a wide range of parameters} even
slightly away from $T_c$ the temperature dependence of $\mu_L$, is
dramatically different from that of the inverse magnetic field
penetration length $\Lambda^{-1}$. }

 Furthermore  because the softest mode with the
mass $\mu_L$ in two band system may be associated with only a
fraction of the total condensate, and because there could be the
second mixed gap mode which can have larger mass $\mu_H$, the
short-range intervortex interaction can be repulsive.
 Since ultimately the sign of the long range
interaction is decided  by the
 competition of $ \Lambda^{-1}$
and $\mu_L$ we plot their temperature dependencies
 in Fig.\ref{Fig:VortexStructure15}(a).
 It shows how in these cases the system
 goes from type-II to type-1.5 behavior
 as temperature is decreased.
 The type-1.5 behavior sets in when
$\mu_L$ becomes smaller than  $ \Lambda^{-1}$,
 and, the density
associated with the light mode is small enough that the system has
a short-range intervortex repulsion.

 To contrast the physics of fundamental
modes in two-band case with singe-band case we plot on
Fig.\ref{Fig:VortexStructure15}(b)
 the product of $\Lambda$ and $\mu_L$.
Note that only {\it infinitesimally}  close to $T_c$, this product
can be interpreted as GL parameter $\kappa$ because the inverse
mass $\sqrt{2}\mu_L^{-1}$ becomes the single component-like GL
coherence length. However away from $T_c$ it represents a mass of
the softest of competing modes and the product $\Lambda \mu_L$ has
a strong and nonmonotonic temperature dependence shown on Fig.
\ref{Fig:VortexStructure15}(b).

\section{Self-consistent calculation of the vortex structure and
non-monotonic vortex interaction energy}

\begin{figure}
\centerline{\includegraphics[width=1.0\linewidth]{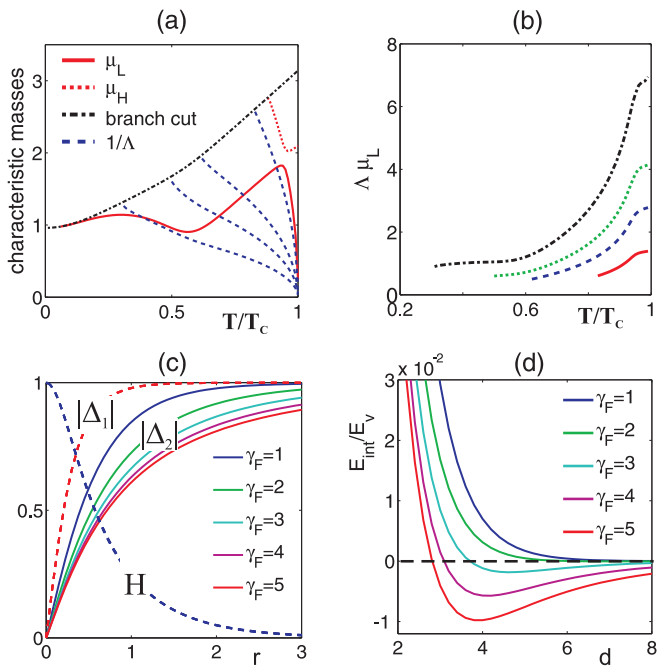}}
\caption{\label{Fig:VortexStructure15} (a) Masses $\mu_{L}$ and
$\mu_{H}$ (red solid and dotted lines) of the composite gap
function fields and inverse London penetration (blue dashed lines)
for the different values of $\Lambda \mu_L
(T_c)/\sqrt{2}=1;2;3;5$. The position of branch cut is shown by
black dash-dotted line. (b) The temperature dependence of the
quantity $\Lambda \mu_L$ for $\Lambda \mu_L(T_c)/\sqrt{2}=1;2;3;5$
(red solid, blue dashed and black dash-dotted lines). (c)
Distributions of magnetic field $H(r)/H(r=0)$, gap functions
$|\Delta_1|(r)/\Delta_{10}$ (dashed lines) and
$|\Delta_2|(r)/\Delta_{20}$ (solid lines) in a single vortex for
the coupling parameters $\lambda_{11}=0.25$, $\lambda_{22}=0.213$
and $\lambda_{21}=0.0025$ and different values of the band
parameter $\gamma_F=1;2;3;4;5$. (d) The energy of interaction
between two vortices normalized to the single vortex energy as
function of the intervortex distance $d$. In panels (c,d) the
temperature is $T=0.6$.
 }
\end{figure}

Next we calculate self-consistently the structure of isolated
vortex for different values of $\gamma_F$. In these calculations
we fix the values of parameters $\sigma_i$ by adjusting the
partial DOS { which in the case of cylindrical Fermi surfaces is
regulated by the ratio of effective masses} so that
$n_2=n_1/\gamma_F$ and $\lambda_{12}=\lambda_{21}/\gamma_F$. We
chose the following values of the coupling parameters
$\lambda_{11}=0.25$, $\lambda_{22}=0.213$. The interband
interaction is small $\lambda_{21}=0.0025$ and the temperature is
$T=0.6$ when $\Delta_{10}\gg\Delta_{20}$. In this case the
composite gap function mode $\tilde{\Delta}_L (r)$ consists mainly
of the weak gap $\bar{\Delta}_2(r)$. Thus, although at the very
long ranges the behavior  of both $|\Delta_1|(r)$ and
$|\Delta_2|(r)$ are determined by the same mass $\mu_L$, the
overall behavior  (i.e. outside asymptotic regimes) of the gap
$|\Delta_1|(r)$ [shown by red dashed lines in
Fig.\ref{Fig:VortexStructure15}(c)] is not very sensitive to the
parameter $\gamma_F$. A complex aspect of the vortex structure in
two-band system is that in general the exponential law of the
asymptotic behavior of the gaps is {\it not} directly related to
the ``core size" at which gaps recover most of
 their  ground state values.  We can characterize this
effect by defining a ``healing" length $L_{\Delta i}$ of the gap
function as follows $|\Delta_i| (L_{\Delta i})= 0.95 \Delta_{i0}$.
Then we obtain that $L_{\Delta 1}\approx 0.8$ for all values of
$\gamma_F$. On the contrary, the healing length $L_{\Delta 2}$ of
changes significantly such that $L_{\Delta 2}= 1.6;\; 2.5;\;
3.2;\; 3.9;\;4.5$ for $\gamma_F=1;\; 2;\; 3;\; 4;\; 5$
correspondingly.

  %%%%%%%%%%%% FIGURE %%%%%%%%%%%%%%%%%%

To demonstrate the type-1.5 behavior we have chosen the parameters
$\sigma_i$ in the self-consistency equation for the current such
that the characteristic magnetic field localization length
$L_H\approx 2$ is much larger than $L_{\Delta 1}$. This leads to a
 existence of regular vortex lattices in a wide range of
strong magnetic fields (i.e. when vortices are closely packed and
thus experience only strong short-range repulsive interaction).
However, the high magnetic field behavior notwithstanding, the
vortex structures shown in Fig.\ref{Fig:VortexStructure15}(c)
clearly shows that $L_{\Delta 1}\ll L_H\ll L_{\Delta 2}$ i.e. the
long-range interaction is attractive and thus the system in fact
belongs to the type-1.5 regime.

Next, to demonstrate the type-1.5 superconductivity i.e.
large-scale attraction and small-scale repulsion of vortices which
originates from disparity of the variations of two gaps,
 we explicitly calculate the intervortex interaction energy.
We evaluate the two-band generalization of the Eilenberger
expression for the free energy of the two vortices positioned at
the points ${\bf r_{R}}=(d/2, 0)$ and ${\bf r_{L}}=(-d/2, 0)$ in
$xy$ plane. Here we generalize to two-band theory the method
developed for calculation of asymptotic vortex interaction in
singe-component theories
%\cite{Leung-VortexAttraction,Kramer,Jacobs-VortexAttraction}.
\cite{Kramer}.
The method assumes that for large separations, in  the region $x<0$
the fields ${\bf H}$, $\Delta_{1,2}$ and $f^{(+)}_{1,2}$
correspond to the single vortex placed at the point ${\bf r_L}$
weakly perturbed by the presence of the second vortex. The
interaction energy can be expressed through the integral over the
line $x=0$ passing
 in the middle between vortices $E_{int}=2\int_{-\infty}^{\infty} d y
 \tilde{E}_{int}(y)$ where
\begin{align}\label{Eq:InteractionEnergy}
 &\tilde{E}_{int}= \int_{-\infty}^{\infty} d y H_v Q_v +  \\ \nonumber
  & T\sum_{j=1,2} \sum_{\omega_n>0}\frac{\sigma_j\Delta_{0j}}{4\omega_n}\int_0^{2\pi} d\theta_p
 \cos\theta_p (f_{Lj} f_{Rj}^+-f^+_{Lj} f_{Rj}).
 \end{align}
 The detailed derivation of the above expression can be found in
 the Appendix\ref{Appendix2}.
 The indices $R(L)$ correspond to the solutions of Eilenberger Eqs.(\ref{Eq:EilenbergerF})
 for isolated vortices positioned at the points ${\bf r_{R(L)}}$.
 The first term in the Eq.(\ref{Eq:InteractionEnergy}) contains
 the  magnetic field $H_v (|{\bf r-\bf r_L}|)$ and the axial component of superfluid
 velocity distribution $Q_v(|{\bf r-\bf r_L}|)$ corresponding to
 the isolated vortex placed at the point ${\bf r}={\bf r_L}$.

In Fig.\ref{Fig:VortexStructure15}(d) the interaction energy
$E_{int}$ is shown as a function of the distance between two
vortices $d$. The energy $E_{int}$ is normalized to the single
vortex energy $E_v$. The plots on
Fig.\ref{Fig:VortexStructure15}(d) clearly demonstrate the
emergence of type-1.5 behavior  when the parameter $\gamma_F$ is
increased. This is manifested in
 the appearance non-monotonic behaviour of $E_{int}(d)$.

\section{Low temperature vortex asymptotics and intrinsic proximity effect.}

  %%%%%%%%%%%% FIGURE %%%%%%%%%%%%%%%%%%
\begin{figure}
\centerline{\includegraphics[width=1.0\linewidth]{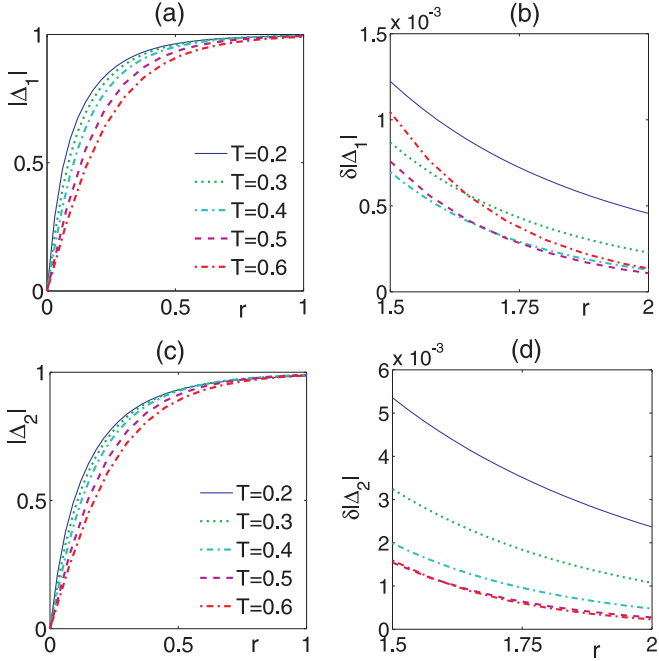}}
\caption{\label{Fig:SoftModeVortexStructure} Gap function profiles
around vortex core for $\lambda_{11}=0.25$, $\lambda_{22}=0.1$,
$\lambda_{12}=\lambda_{21}=0.05$.  (a)  and (c): Variation of the
gap functions $|\Delta_{j}|(r)/\Delta_{j0}$ ($j=1,2$) near the
core. (b)  and (d): The behaviour of gap function deviations from
the vacuum state
$\delta\Delta_{j}(r)=1-|\Delta_{j}|(r)/\Delta_{j0}$ at longer
range.  Note that in this temperature span the
  {\it higher} is the temperature the {\it faster} is the long distance decay of
  $\delta\Delta_{j}(r)$, which reflects the found fact in two-band
system the field mass can ibcrease with raising temperature [see
also Fig.\ref{Fig:Asymptotics1}(d).] }
\end{figure}

Finally we discuss the two-band superconductor with
$\Delta_{20}\ll\Delta_{10}$ at $T\rightarrow 0$. Note that
qualitatively similar regime is
  realized in the two-band superconductor $MgB_2$ \cite{gurevich}. To model such situation we choose the coupling
  constants $\lambda_{11}=0.25$, $\lambda_{12}=\lambda_{21}=\lambda_{J}=0.05$ and
 consider various values of $\lambda_{22}$.
The temperature dependencies of the mass $\mu_L(T)$
 for different values of $\lambda_{22}$ are shown in
 the Fig.\ref{Fig:Asymptotics1}(d).
Note that in this case, decreasing
of intraband coupling $\lambda_{22}$ leads to the
decreasing of the $\mu_L$ at low temperatures.
This anomalous behaviour of the characteristic length scale is clearly manifested in the vortex structure
 shown in Fig.(\ref{Fig:SoftModeVortexStructure}).
   The near-core gap function profiles
   [Fig.\ref{Fig:SoftModeVortexStructure}(a,c)]
 feature shrinkage  of the vortex core at
  decreasing temperature, similarly to clean single-band
  superconductors\cite{KramerPesch}. However the asymptotics of gap
  functions   [Fig.\ref{Fig:SoftModeVortexStructure}(b,d)] are
  drastically different from the single-band case. Indeed, it can
  be seen that  { in a certain temperature domain} the lower the temperature, {the slower is the
 recovery the gap functions at large distances from the core}.
  Such behavior in the two-band system is clearly in a sharp contrast with the overall vortex
    core shrinking with decreasing temperature in clean single-band superconductors.

 Note that in the above case, at low temperatures
we have  $\mu_L\approx 2\sqrt{\Delta_{20}^2+(\pi
 T)^2}/v_{F2}$. For the especially interesting { regime} of purely interband proximity effect-induced superconductivity in the weak band
 we can consider the limit $T\gg \Delta_{20}/\pi$. Then $\mu_L\approx
 \xi^{-1}_N$, where $\xi_N=v_{F2}/(2\pi T)$ is the coherence length in a pure
 normal metal \cite{deGennes} describing the penetration length of superconducting
 correlations induced by the proximity effect in superconductor/normal metal (SN) hybrid structures\cite{proximity}.
 Thus we obtain that the intrinsic proximity effect due to the interband
 coupling\cite{GL2mass2} can in certain cases be described by the similar length scale as
 the usual one in SN hybrid structures. At the temperature interval $\Delta_{20} \ll \pi T\ll \Delta_{10}$
 the mass $\mu_L(T)$ grows linearly with temperature
 [Fig.\ref{Fig:Asymptotics1}(d)].
 \section{Conclusion}

In conclusion, the rapidly growing family of discovered multiband
superconductors ($MgB_2$, Iron pnictides etc) requires
understanding and classification of possible magnetic response of
systems with multiple superconducting gaps. Here we reported a
microscopic theory of magnetic response  of a superconductor with
two bands (the developed approach can be generalized to the case
of a higher number of bands).
%We have shown that
%the GL parameter $\kappa$ is well defined only extremely close to $T_c$
%and in general
%does not determine the magnetic response of the system
%if the coupling between bands is not very strong.
We have shown that new physics which arises in multiband systems
is the existence of several {\it mixed} gaps modes.
This, in a range of parameters results in the existence of the type-1.5
superconducting regime.
%At elevated temperature the presented here
 %theory verifies and gives the microscopic interpretation of
%the type-1.5 superconductivity which was suggested in the framework of
% phenomenological GL models
%\cite{GL2mass,GL2mass2}. Next
We described the system properties and emergence of type-1.5
regimes in the entire temperature regimes, in particular beyond
the validity of  a two-component GL theory. The universal feature
of all the regimes supporting type-1.5 behavior is the
thermodynamic stability of vortex excitations in spite of the
existence of a mode which varies at a fundamental length scale
larger than the magnetic field penetration length.
 It results in  non-monotonic
vortex interaction and appearance of the additional Semi-Meissner
phase  in low magnetic fields which is  a macroscopic phase
separation into (i) domains of two-component vortex state and (ii)
vortex clusters where one of the components is suppressed.

\section{Acknowledgments}
%We thank D. Agterberg, A. Gurevich, A. Mel'nikov and J. Sauls for
%stimulating discussions.
The work is supported by the NSF CAREER Award No. DMR-0955902, the
Knut and Alice Wallenberg Foundation through the Royal Swedish
Academy of Sciences and by the Swedish Research Council,
``Dynasty" Foundation, Presidential RSS Council (Grant No.
MK-4211.2011.2) and Russian Foundation for Basic Research.\

\appendix
\section{Asymptotical behaviour of the gap functions.}
\label{Appendix1}

We focus on the structure of the isolated axially symmetric vortex
in two-band superconductor characterized by the non-trivial phase
winding of
 the gap functions:
 \begin{equation}\label{App1Eq:VortexOP}
\Delta_{1,2}=|\Delta_{1,2}|(r)e^{i\varphi}.
\end{equation}
 We begin by considering the asymptotical behaviour of the gap
 functions at distances far from the vortex core when the deviations of all fields from
the homogeneous values are small. In this case the Eilenderger
Eqs.(\ref{Eq:EilenbergerF}) can be linearized in order to find the
asymptotical behavior of the gap functions modules
$|\Delta_{1,2}|(r)$. { To  compare with the different
linearlization problem in  single-band case see
Ref.(\onlinecite{Eilenberger-Buttner}).}

 To determine the asymptotic behaviour we
   use the transformation $f\rightarrow
   f e^{i\varphi}$, $f^+\rightarrow  f^+e^{-i\varphi}$ and
 rewrite the Eilenberger Eqs.(\ref{Eq:EilenbergerF})
   in terms of the deviations from the vacuum state values
$\bar{\Delta}_j=\Delta_{j0}-|\Delta_j|$ and
 $\bar{f}_j={f}_{j0}-{f}_j$,
  $\bar{f}^{+}_j={f}^{+}_{j0}-{f}^{+}_j$.
 Then keeping the first order terms $\bar{f}_{\Sigma(d)}$ and $\bar{\Delta}_j$ in the
l.h.s. we can rewrite the Eilenberger Eqs. in the following form
  (we omit the band index for brevity):
 \begin{align}\label{App1Eq:EilenbergerF-lin}
  &v_F{\bf n}\nabla \bar{f}_{\Sigma}+2\omega_n
  \bar{f}_{d}=X_\Sigma \\ \nonumber
  &v_F{\bf n} \nabla \bar{f}_{d}+2\frac{\Omega_n^2}{\omega_n} \bar{f}_{\Sigma}-i\frac{2\Delta_0}{\Omega_n}{\bf n}{\bf Q}
  -\frac{4\omega_n}{\Omega_n}\bar{\Delta}=X_d.
 \end{align}
 where the higher order terms in $\bar{\Delta}_j$, $\bar{f}$ and $\bar{f}^+$
  are incorporated in the r.h.s. functions $X_{\Sigma (d)}=X_{\Sigma (d)}({\bf n_p},\omega_n, {\bf
 r})$. In Eqs.(\ref{App1Eq:EilenbergerF-lin}) we introduce $\Omega_{n}=\sqrt{\omega_n^2+\Delta_{0}^2}$
 and the functions $\bar{f}_{\Sigma}=\bar{f}+\bar{f}^+$ and $\bar{f}_{d}=\bar{f}-\bar{f}^+$. The higher order terms are incorporated in the
 functions $X_{\Sigma (d)}=X_{\Sigma (d)}({\bf n_p},\omega_n, {\bf
 r})$.

  Then we take the real part of the Eqs.(\ref{App1Eq:EilenbergerF-lin})
 to obtain the following system
 \begin{align}\label{App1Eq:EilenbergerF-lin-R}
 &v_F{\bf n_p}\nabla {f}^r_{\Sigma}+2\omega_n
 {f}^r_{d}=X^r_\Sigma \\ \nonumber
 & v_F{\bf n_p} \nabla {f}^r_{d}+2\frac{\Omega_n^2}{\omega_n}
 {f}^r_{\Sigma}-
 \frac{4\omega_n}{\Omega_n}\bar{\Delta}=X^r_d.
 \end{align}
 Here omit the band index for brevity and denote $f^r_{\Sigma(d)}=Re
 \bar{f}_{\Sigma(d)}$.
  Below we will
  find the asymptotic of the gap fields treating the
  nonlinear terms in the r.h.s. of
  Eqs.(\ref{App1Eq:EilenbergerF-lin-R}) as source functions.

The solution of Eqs.(\ref{App1Eq:EilenbergerF-lin-R}) can be found
in the momentum representation $f^r_{\Sigma,d} ({\bf k})=\int
f^r_{\Sigma,d}({\bf r}) \exp(-i {\bf k r}) d^2 {\bf r}$.
 Then we get
\begin{equation}\label{App1Eq:f-r}
{f}^r_\Sigma=\frac{\omega_n^2}{\Omega_n}\frac{8\bar{\Delta}}{4\Omega_n^2+({\bf
{\bf v_{F}} k})^2}+ M({\bf v_{F}}{\bf k},\omega_n)
 \end{equation}
 where the last term incorporates the higher order corrections:
\begin{equation}\label{App1Eq:Mnonlinear}
M({\bf v_{F} k},\omega_n)=\frac{2\omega_n}{4\Omega_n^2+({\bf v_{F}
k})^2} \left(X_d^r-\frac{i {\bf v_{F}k}}{2\omega_n}
X_\Sigma^r\right).
\end{equation}

 After substituting it to the self-consistency
Eq.(\ref{Eq:SelfConsistencyOP}) we get the expression for the
order parameter
 \begin{equation}\label{App1Eq:OpFourier}
 \bar{\Delta}_i(k)= \hat R^{-1}_{ij} N_j(k)
 \end{equation}
 where
\begin{equation}\label{App1Eq:Nj}
N_i(k)= \frac{\lambda_{ij} T}{2} \sum_{n=0}^{N_d} \int_0^{2\pi}
M_j d\theta_p
\end{equation}
 and the elements of the matrix $\hat R=\hat R(k)$ are defined by
 $R_{ii}=(\lambda_{ii} S_i-1)$ and $R_{ij}=\lambda_{ij} S_j$, where
\begin{equation}\label{App1Eq:Sj-1}
S_j= 4T\sum_{n=0}^{N_d}\frac{\omega_n^2}{\Omega_{nj}}\int_0^{2\pi}
\frac{d\theta_p}{4\Omega_{nj}^2+({\bf v_{Fj} k})^2}.
\end{equation}
The integrals entering the expressions (\ref{App1Eq:Sj-1}) above
are
 $$
 \int_0^{2\pi}
  \frac{d\theta_p}{b^2+(\sin\theta_p)^2}=\frac{2\pi}{b\sqrt{b^2+1}}
 $$
 so that
 \begin{equation}\label{App1Eq:Sj}
 S_j(k)=4\pi T\sum_{n=0}^{N_d} \frac{\omega_n^2}{\Omega_{nj}^2}\frac{1}{\sqrt{4\Omega_{nj}^2+(v_{Fj} k)^2}}.
 \end{equation}
  The source functions $N_j(k)$ come from
the nonlinear terms $X^r_{\Sigma,d}$  in Eilenberger
Eqs.(\ref{App1Eq:EilenbergerF-lin-R}).

 { The Eq.(\ref{App1Eq:OpFourier})
 is the two-band  response function.
 To compare with the single-band response function see \cite{Leung-Jacobs}.}
 In general the
 real space asymptotic behaviour of the order parameter (\ref{App1Eq:OpFourier})
 is determined by the contributions of the singularities of the response function
 $\hat R^{-1}(k)$ which are poles and
  branch points at $k=2i\Omega_{nj}/v_{Fj}$. Analogously to the consideration in
 Ref.(\onlinecite{Leung-Jacobs}) we assume the branch
 cuts to lie along the imaginary axis from $k=2i\Omega_{nj}/v_{Fj}$ to
 $k=i\infty$. The poles are determined by the zeros of the determinant
 $D_R(k)={\rm Det} \hat R(k)=0$, so that
$$
 D_R(k)=(1-\lambda_{11} S_1)(1-\lambda_{22}
 S_2)-\lambda_{12}\lambda_{21}S_1S_2.
  $$

 Since we are interested in the asymptotic behaviour of the
 order parameter, we need only to take into account the poles of
 $\hat R^{-1}(k)$ lying in the upper complex half plane below all the
 branch cuts. In this case all the
 zeros of the function $D_R (k)$ are purely imaginary $k^*=iq_n$.
 Each of them can be associated with the particular
 mass of the gap function field $\mu_n=1/q_n$
 which determine the characteristic length scale of the gap
 function variation. On the other hand the contribution from the
 branch cut contains all the length scales which are larger than
 the threshold one given by position of the lowest branch point
 $k=i q_{bp}$ where
\begin{equation}\label{App1Eq:BranchCut}
q_{bp}=2\min(\Omega_{02}/v_{F2}, \Omega_{01}/v_{F1}).
\end{equation}

 \section{Energy of interaction between two vortices}
 \label{Appendix2}
  \subsection{General free energy expression}
  The two-band generalization of the Eilenberger expression for the free
energy\cite{Eilenberger:FreeEnergy} reads as follows
\begin{align}\label{App2Eq:FreeEnegry}
 &F({\bf r})=
 \frac{{\bf H^2}}{2}+\tilde{\rho}_{11}
 |\Delta_1|^2+ \tilde{\rho}_{22}
 |\Delta_2|^2 +
 \\ \nonumber
 &\tilde{\rho}_{J}\left(\Delta_1\Delta_2^*+\Delta_2\Delta_1^*\right)
  + F_{I1}+F_{I2}
 \end{align}

 where
 $$
 \begin{pmatrix}
   \tilde{\rho}_{11} & \tilde{\rho}_{12} \\
   \tilde{\rho}_{21} & \tilde{\rho}_{22} \
 \end{pmatrix}=
 \frac{1}{\kappa^2}\begin{pmatrix}
   \rho_{11} & \rho_{12} \\
   \rho_{21} & \rho_{22} \
 \end{pmatrix}^{-1}
 $$
$\tilde{\rho}_J=\tilde{\rho}_{12}=\tilde{\rho}_{21}$ and
 $$
 F_{Ij} =-  \frac{T}{\kappa^2} \sum_{\omega_n>0} \int_0^{2\pi} n_j I_j(\omega_n,\theta_p, {\bf r})
 d\theta_p
 $$
 with
\begin{align}\label{App2Eq:EnegryI}
 &I_j(\omega_n,\theta_p, {\bf r})= \Delta_j^* f_j+\Delta_j f_j^+\\ \nonumber
 & +(g_j-1)\left[2\tilde{\omega}_n + \frac{v_{Fj}}{2}{\bf n_p \nabla} \left(\ln f_j-\ln f_j^+\right)\right]
 \end{align}
 where $j=1,2$ and
 $$
 \tilde{\omega}_n=\omega_n+iv_{Fj}{\bf n_p A}/2.
 $$
 Then the variation of the free energy (\ref{App2Eq:FreeEnegry})
 with respect to the fields ${\bf A}$ and $\Delta$ gives the
 self-consistency Eqs.(\ref{Eq:SelfConsistencyCurrent}) and (\ref{Eq:SelfConsistencyOP}) correspondingly. The
 variation over $f$ and $f^+$ with the normalization condition
 taken into account yields the Eilenberger Eqs.(\ref{Eq:EilenbergerF}).
 Provided the functions $f,f^+,g$ satisfy the Eqs.(\ref{Eq:EilenbergerF}) the
 expression (\ref{App2Eq:EnegryI}) can be rewritten as
 \begin{equation}\label{App2Eq:EnegryI1}
 I_j(\omega_n,\theta_p,{\bf r})= \frac{\Delta_j^* f_j+\Delta_j f_j^+}{1+g_j}.
 \end{equation}

  \subsection{Linearized theory of vortex interaction}

{ To calculate the energy of vortex interaction we evaluate the
 free energy expression for the system of two vortices positioned
at the points ${\bf r_{R}}=(d/2, 0)$ and ${\bf r_{L}}=(-d/2, 0)$
in $xy$ plane.  Here we employ the method
similar to that in \cite{Kramer}. %,\cite{Jacobs-VortexAttraction}, \cite{Leung-VortexAttraction}.}

Let us consider the half-plane $x<0$ containing only one of the
vortices. { { We decompose the gap function into amplitude and
phase (we omit the band index for brevity)
\begin{equation}\label{App2Eq:decompose}
\Delta({\bf r})=|\Delta|({\bf r})\exp(i\Phi).
\end{equation}
 The total phase can be written in the following form
 $\Phi=\Phi_L+\Phi_R+\Phi_{ns}$, where
 $$
 \Phi_{L(R)}({\bf r})=\arctan\left(
 \frac{y-y_{R(L)}}{x-x_{R(L)}}\right)
 $$
 are the vortex phases and $\Phi_{ns}({\bf r})$ is a regular part of the
 phase.}} At the region $x<0$ we can make the gauge transformation
 removing the phase $\Phi_{R}({\bf r})$, since it does not contain
 singularities. After this transformation we can
  assume that the fields ${\bf A}$, $\Delta_{1,2}$ and $f^{(+)}_{1,2}$
 correspond to the solutions for a single vortex placed at the point ${\bf r_L}$
 weakly perturbed by the presence of the second vortex.
 $$
 {\bf A} = {\bf A_{v}} + \delta {\bf Q}; \;\;\; \Delta_j = \Delta_{vj} +
 \delta\Delta_j
 $$
 $$
 f_j = f_{vj} + \delta f_j;\;\;\; f^+_j = f^+_{vj} + \delta f^+_j.
 $$
 where we have introduced the  superfluid velocity induced by the second
 vortex $\delta{\bf Q}={\bf A_{R}}-\nabla \Phi_R$. Then we obtain
\begin{eqnarray}\label{App2Eq:deltaI}
  &\delta I_j=  (\delta \Delta_j f_{vj}^+ + \delta \Delta_j^* f_{vj})+  (\Delta_{vj}\delta f^+_j+\Delta_{vj}^*\delta
 f_j)
  \\ \nonumber
  &+ iv_{Fj} (g_{vj}-1) {\bf n_p \delta Q} + 2\tilde{\omega}_n \delta
 g_j \\ \nonumber
 &+v_{Fj} \frac{\delta g_j }{2}{\bf n_p \nabla} \left(\ln f_{vj}- \ln f^+_{vj}\right)\\ \nonumber
 &+  v_{Fj} \frac{(g_{vj}-1)}{2}{\bf n_p \nabla} \left(\frac{\delta f_j}{f_{vj}}- \frac{\delta f^+_j}{f^+_{vj}}\right)
 \end{eqnarray}
 where
 $$
 \delta g_j= -(f_{vj}\delta f^+_j + f_{vj}^+\delta f_j)/2g_{vj}.
 $$
 The last two terms in Eq.(\ref{App2Eq:deltaI}) can be rewritten as
 follows
 \begin{align}
 &\frac{1}{2g_v} \left[ \delta f ({\bf n_p \nabla}) f_v^+ - \delta
 f^+ ({\bf n_p\nabla})f_v \right]\\ \nonumber
 &\frac{({\bf n_p \nabla})}{2}\left[(g_v-1)\left(\frac{\delta f}{f_v}- \frac{\delta
 f^+}{f^+_v}\right)\right].\nonumber
 \end{align}
 The first term in this expression cancels with the
 second and forth terms in Eq.(\ref{App2Eq:deltaI}).
 For the variation of magnetic field energy in
 Eq.(\ref{App2Eq:FreeEnegry}) we obtain
 $$
  {\bf H_v} \delta{\bf H} =
  \nabla\cdot (\delta{\bf Q}\times{\bf H_v})+
  \nabla\times {\bf H_v} \cdot \delta{\bf Q}.
 $$
 Then we are left
 with the non-zero terms
 \begin{align}
 &\delta F= \nabla \cdot    \delta{\bf Q}\times{\bf H_v}  \\
 \nonumber
 &-\frac{T}{2\kappa^{2}}\sum_{j,\omega_n}  n_jv_{Fj}\int_0^{2\pi}
 d\theta_p \nabla \cdot{\bf n_p}\left[(g_{vj}-1)\left(\frac{\delta f_j}{f_{vj}}-
  \frac{\delta f^+_j}{f^+_{vj}}\right)\right]
 \end{align}

 The energy of vortex interaction is  $ E_{int}= 2\int \delta F d {\bf
 r}$. It can be expressed through the integral over the line $x=0$
 so that $E_{int}=2\int_{-\infty}^{\infty } d y {\bf x}\cdot {\bf e}_{int} $
 \begin{align}\label{App2Eq:InteractionEnergy-1}
 &{\bf e}_{int}= \delta{\bf Q}\times{\bf H_v} - \\ \nonumber
 & \frac{T}{2\kappa^{2}} \sum_{j,\omega_n}
 n_jv_{Fj}\int_0^{2\pi} d\theta_p {\bf n_p}
 \left[(g_{vj}-1)\left(\frac{\delta f_j}{f_{vj}}- \frac{\delta f^+_j}{f^+_{vj}}\right)
  \right] .
 \end{align}
 {{ To evaluate the second term in
 Eq.(\ref{App2Eq:InteractionEnergy-1}) it is convenient to
 bring the Eqs.(\ref{Eq:EilenbergerF}) to the gauge invariant
 form\cite{DopplerShift}
 decomposing the gap functions into amplitude and
phase (\ref{App2Eq:decompose}) and transforming the Green's
functions as $f\rightarrow f e^{i\Phi}$, $f^+\rightarrow
f^+e^{-i\Phi}$.
 Then at the line $x=0$ we can put}}
 $$
 f_{vj}=f_{0j}+ f_{Lj}; \;\;\;f^+_{vj}=f_{0j}+ f^+_{Lj},
 $$
 where
 $f_{0j}=\Delta_{0j}/\sqrt{\Delta_{0j}^2+\omega_n^2}$.
 Also we denote $\delta f_j= f_{Rj}$, $\delta f^+_j= f^+_{Rj}$ [$L(R)$ stand for left (right) vortices].
 Therefore up to the second  order terms we obtain
\begin{align}\label{App2Eq:1}
&(g_{vj}-1)\left(\frac{\delta f_j}{f_{vj}}-
 \frac{\delta f^+_j}{f^+_{vj}}\right)=\frac{g_{0j}-1}{f_{0j}} \left(f_{Rj}-f_{Rj}^+\right)\\ \nonumber
 & - \frac{1}{2g_{0j}}\left(f_{Lj}+f^+_{Lj}\right)\left(f_{Rj}-f^+_{Rj}\right) \\ \nonumber
 &+ \frac{g_{0j}-1}{f^2_{0j}} \left(f^+_{Rj} f^+_{Lj}-f_{Rj} f_{Lj}\right).
 \end{align}
 Now we use the symmetry relations $f_{L,R}(n_x,n_y)=f^*_{R,L}(-n_x,n_y)$
 and $f^*(-{\bf n_p})=f^+({\bf n_p})$.
 Then the contribution to the interaction energy (\ref{App2Eq:InteractionEnergy-1})
 from the first order term in Eq.(\ref{App2Eq:1})
 cancels with the analogous contribution from the left vortex.
 Also from the symmetry relations we obtain
 \begin{align}
 &Re \int_0^{2\pi} \cos \theta_p f_L f_R d\theta_p =0 \\ \nonumber
 &Re \int_0^{2\pi} \cos \theta_p f^+_L f^+_R d\theta_p =0.
 \end{align}
 On the other hand
 $$
 Im \int_0^{2\pi} \cos \theta_p \left(f_L f_R - f^+_L f^+_R\right) d\theta_p
 =0.
 $$
 { Therefore we get for the interaction energy}
 $$E_{int}=2\int_{-\infty}^{\infty} d y
 \tilde{E}_{int}(y),$$
 where
\begin{align}\label{App2Eq:InteractionEnergy}
 &\tilde{E}_{int}= H_v Q_v + \\ \nonumber
 &T\sum_{j=1,2}\sigma_j \sum_{\omega_n>0}
 \frac{\Delta_{0j}}{4\omega_n}\int_0^{2\pi} d\theta_p
 \cos\theta_p (f_{Lj} f_{Rj}^+-f^+_{Lj} f_{Rj}),
\end{align}
and $\sigma_j=\kappa^{-2}  n_jv_{Fj}$.

\end{document}